\documentclass[preprint,12pt]{elsarticle}

\usepackage{algorithmic}
\usepackage{algorithm}

\usepackage{amssymb}

\journal{Astronomy and Computing}

\begin{document}

\newcommand{\R}{{\rm I}\!{\rm R}}
\newcommand{\Co}{{\rm I}\!\!\!\!{\rm C}}

\begin{frontmatter}



\title{{\em{DESAT}}: an SSW tool for {\em{SDO/AIA}} image de-saturation}


\author[pippo1]{Richard A Schwartz}
\author[pippo2]{Gabriele Torre}
\author[pippo3]{Anna Maria Massone}
\author[pippo2,pippo3]{Michele Piana}

\address[pippo1]{NASA Goddard Space Flight Center and Catholic University of America, Greenbelt, MD 20771, USA }
\address[pippo2]{Dipartimento di Matematica, Universit\`a di Genova, Genova, Italy}
\address[pippo3]{CNR - SPIN, Genova, Italy}

\begin{abstract}
Saturation affects a significant rate of images recorded by the {\em{Atmospheric Imaging Assembly}} on the {\em{Solar Dynamics Observatory}}. This paper describes a computational method and a technological pipeline for the de-saturation of such images, based on several mathematical ingredients like Expectation Maximization, image correlation and interpolation. An analysis of the computational properties and demands of the pipeline, together with an assessment of its reliability are performed against a set of data recorded from the Feburary 25 2014 flaring event.
\end{abstract}


\end{frontmatter}


\section{Introduction}
\label{introduction}
As in any optical system, telescopes introduce blur into the images they produce. The characteristics of such a blur are encoded in the instrument point spread function (PSF), which describes the response of the whole optical hardware to the signal produced by a distant point source. Deconvolving any blurring and artifacts associated with this PSF is a crucial step among the variety of tasks to be addressed in image processing. Unlike many instruments, whose PSF is made of just a main central peak associated with signal diffusion, some telescopes have a PSF with a double structure, whereby a second, more complex component overlays the diffusion core, containing peaks that replicate the central one according to regular patterns of varying intensity \cite{beba78}. Such replications are caused by diffraction which is produced by the waves scattering off the regular wire meshes that support the instrument entrance filters. 

During intense solar flares the diffraction fringes appear as artefacts at long range from a central core but these are a timely source of information on the true image as the direct information has been destroyed by saturation effects in the recording hardware. In fact, the principle of CCD devices \cite{makl97} is to convert photons into electrons. The electrons that accumulate in the CCD wells are read out and converted to data numbers (DNs) by software. The amount of charge that a single pixel can accumulate is finite and coarsely related to its area. However, the probability of trapping an electron within a pixel well decreases when the well is approaching saturation, which implies that the linear relation between light intensity and signal degrades and the response of a saturated pixel drops. For intense photon fluxes, the major fraction of the signal accumulates in the CCD image pixels only up to the saturation limit while another part is coherently scattered by the support to produce the diffraction fringes in the remote pixels without any deviation from a linear response. Therefore, the diffraction signal is unaffected by saturation; further, the diffraction pattern is a direct signature of the photon flux in the saturated region and therefore, in principle, an inverse process can restore the original scene. 

Image saturation has been an issue for a number of telescopes in solar astronomy, like the NASA {\em{Solar Terrestrial Relations Observatory (STEREO)}} \cite{soetal00} and the NASA {\em{Transition Region and Coronal Explorer (TRACE)}} \cite{gbetal06}. However, with the introduction of the {\em{Atmospheric Imaging Assembly (AIA)}} in the {\em{Solar Dynamics Observatory (SDO)}} \cite{leetal12} de-saturation has become a big data issue. Indeed {\em{AIA}} provides full-disk $4096 \times 4096$ pixel images of the solar chromosphere and corona in seven EUV wavelengths with a time cadence of $12$ seconds and saturation affects a large fraction of these images: a coarse estimate, based on the fraction of saturated frames in typical C-class, M-class and X-class events and on the typical number of such events provided by the {\em{RHESSI}} flare-list, suggests that more than $10^5$ {\em{AIA}} images per year are saturated. This pathology prevents a systematic exploitation of {\em{AIA}} in both the accomplishment of its scientific objectives and the integration with data from other observation missions. 

{\em{DESAT}} is a software package now available in {\em{Solar SoftWare (SSW)}} that realizes an automatic de-saturation of {\em{AIA}} images by using a correlation/inversion analysis of the diffraction fringes produced in the telescope observations \cite{sctopi14}. The conceptual features of {\em{DESAT}} are essentially two. First, the tool strongly relies on the shape and characteristics of the instrument PSF in both its components, the diffusion core and the diffraction PSF. Second, the mathematical content at the basis of DESAT is not at all trivial and involves Expectation Maximization (EM) \cite{shva82} for realizing image segmentation and "inverse" diffraction. Moreover, this same mathematical framework makes {\em{DESAT}} easily applicable to image de-saturation in the case of instruments other then {\em{SDO/AIA}}, under the condition that these instruments are characterized by a similar two-component PSF. 

{\em{DESAT}} pipeline also provides a module for reducing secondary effects typical of instruments like {\em{AIA}}. When a pixel reaches saturation, additional charge cannot be accommodated in the pixel well and therefore spills over into adjacent pixels, often causing them to saturate. This spread of charge to a pixel region around the primary saturation core is named {\em{blooming}} and typically appears as bright ray along a preferential direction depending on the CCD design as well as saturated pixels surrounding the primary saturation. {\em{DESAT}} takes care of blooming and ameliorates its effects by exploiting the fact that each saturated {\em{AIA}} image is embedded in a time sequence of variable duration, made of acquisitions performed with smaller exposure times and therefore non-saturated. {\em{DESAT}} utilizes the information contained in the sequence of non-saturated images to interpolate an estimate of the pixel content in the blooming region. Although the applications considered so far show that this approach is rather reliable, we note that it is significantly less robust than the segment of {\em{DESAT}} pipeline reducing primary saturation: the routines of such segment, in fact, utilize information provided by the instrument at the right time and directly related to primary saturation by the physics of diffraction.

The plan of the paper is as follows. Section 2 provides the mathematical setup for {\em{DESAT}}. Section 3 describes the pipeline and Section 4 performs some numerical tests to validate the effectiveness of the toolbox. Our conclusions are offered in Section 5, together with a discussion of the main open issues.



\section{Mathematical setup}
\begin{figure}
\begin{center}
\begin{tabular}{c}
\includegraphics[width=11.cm]{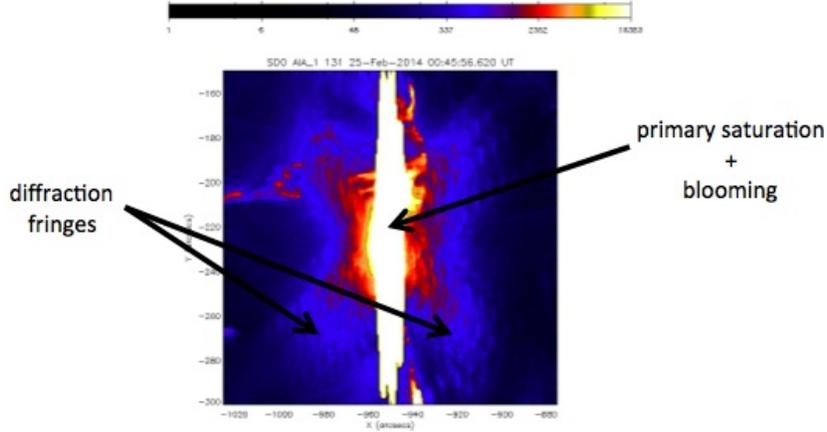} 
\end{tabular}
\caption{An example of saturated {\em{AIA}} image with highlighted the overall (primary + blooming) saturation region and the diffraction fringes. The event is the one occurred on February 25 2014, time point 00:45:56 UT.}
\label{fig:regions}
\end{center}
\end{figure}

This Section briefly summarizes the mathematical theory at the basis of the de-saturation approach described in much more details in \cite{toetal15}. The starting point for this theory is that the inverse diffraction problem at the basis of the {\em{DESAT}} toolbox can be described by the linear model
\begin{equation}\label{model-equation}
g = Kf + b~,
\end{equation}
where (see Figure \ref{fig:regions}) $g$ is the recorded signal in correspondence of the diffraction fringes, $f$ is the true flux in the primary saturation region, $K$ is the diffraction component of the {\em{AIA}} PSF mapping the primary saturation region onto the diffraction pattern and $b$ is the portion of the image background associated to the diffraction fringes. 

Then the mathematical challenge of solving equation (\ref{model-equation}) is many-fold and requires the following steps:
\begin{description}
\item{{\bf{Segmentation.}}} This step separates the image to de-saturate in four regions: the one containing the diffraction fringes, the primary saturation region, the blooming region and the remaining part of the image, which will stay untouched during the overall de-saturation process.  We first of all observe that the diffraction pattern and the primary saturation region are directly connected, since the diffraction fringes are significantly illuminated, via the diffraction PSF, just by the flux directed toward the pixels contained in the primary saturation region. The crucial issue of this step is therefore to segment the saturation and the blooming regions. In fact, at a first inspection of the image, such regions cannot be distinguished and are made by those pixels whose grey level content is bigger than 16,000 DN pixel$^{-1}$, i.e. the saturation threshold for {\em{AIA}}. However, the diffraction PSF strongly correlates just with the pixel in the primary saturation, while the correlation between $K$ and the blooming part of the region must be very weak. Therefore {\em{DESAT}} computes the correlation map by means of one iteration of the EM algorithm and retains as primary saturation pixels just those pixels for which such map (after projection onto the data space by means of the diffusion PSF) has components larger than the saturation value.
\item{{\bf{Reconstruction.}}} This step solves equation (\ref{model-equation}) by means of EM. Assuming that the diffraction fringes $g$ are affected by Poisson noise (CCD data are more correctly quasi-Poisson but the approximation is reasonable in this context), EM implements maximum likelihood under positivity constraint by optimally stopping the iteration \cite{shva82}
\begin{equation}\label{EM}
f^{k+1} = \frac{f^k}{K^T {\bf{1}}} K^T \left( \frac{g}{Kf^k + b} \right)~,
\end{equation}
where ${\bf{1}}$ is made of all ones. Several different stopping rules can be applied for EM. We empirically found that the one mathematically discussed in \cite{bepi14} and applied in \cite{beetal13} for image reconstruction from count modulation profiles recorded by the {\em{Reuven Ramaty High Energy Solar Spectroscopic Imager (RHESSI)}} works very appropriately also in the case of {\em{AIA}} data and therefore we used it in {\em{DESAT}}.
\item{{\bf{Image synthesis.}}} The map provided by EM in the reconstruction step is an estimate of the incoming flux in correspondence to the primary saturation region. Since saturation occurs in the data region, DESAT projects the EM solution onto the data space by means of the core diffusion PSF and then synthesize the de-saturated image by glueing the result of the projection onto the image background.
\end{description}

All three steps described above exploit the knowledge of an estimate of the image background. {\em{DESAT}} constructs this background map via interpolation of the pixel content in non-saturated images acquired in a time range containing the time point when the saturated image has been recorded. Once the background map is available different sub-regions of it are used by {\em{DESAT}} for different purposes: the sub-region corresponding to the diffraction fringes is used in the iterative solution of equation (\ref{model-equation}) by means of EM (it is matrix $b$ in (\ref{model-equation}) and (\ref{EM})); the subregion where the overall saturation region (primary saturation + blooming) is projected by diffraction is used to compute correlation in the segmentation process; the subregion corresponding to the blooming region is used in the synthesis of the final de-saturated image.

\section{The {\em{DESAT}} pipeline}
\begin{figure}
\begin{center}
\begin{tabular}{c}
\includegraphics[width=9.cm]{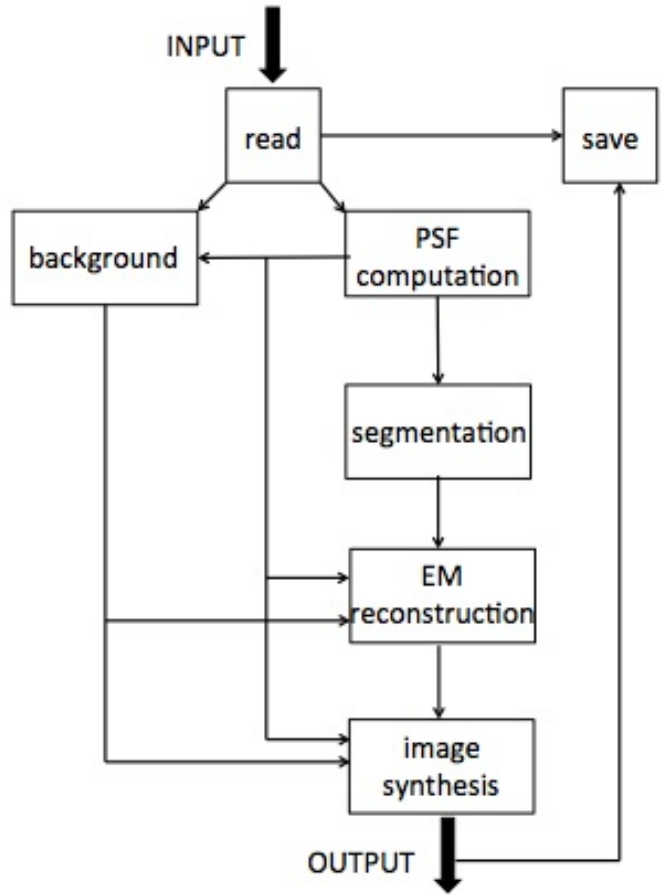} 
\end{tabular}
\caption{The DESAT pipeline.}
\label{fig:pipeline}
\end{center}
\end{figure}

The pipeline realized by {\em{DESAT}} in {\em{Interactive Data Language (IDL)}} and currently available in SSW is illustrated in Figure \ref{fig:pipeline} and is made of the following steps:
\begin{description}
\item{{\bf{INPUT:}}} this step loads a file containing a set of saturated and unsaturated {\em{AIA}} images, corresponding to a selected time interval and a vector indicating the specific wavelengths to process. These two ingredients are saved in the 'save' step.
\item{{\bf{read:}}} this step simply reads the files in the datasets associated to the selected wavelengths.
\item{{\bf{background estimation:}}} this is one of the computational cores of the pipeline. It provides an estimate of the background for the whole field of view covered by the image.
\item{{\bf{PSF computation:}}} this step computes the diffusion PSF component, the diffraction PSF component and the global PSF.
\item{{\bf{segmentation:}}} this step utilizes correlation to identify the position of primary saturation pixels, of blooming pixels and of the pixels in the diffraction fringes.
\item{{\bf{EM reconstruction:}}} this step utilizes EM to reconstruct the incoming EUV flux in correspondence of the primary saturation region.
\item{{\bf{image synthesis:}}} this step projects the reconstructed flux in the image space by means of the diffusion PSF and glues the result on the background to provide the unsaturated image.
\item{{\bf{OUTPUT:}}} the output of the pipeline is made of a dataset of de-saturated {\em{AIA}} images which are saved again in the 'save' step.
\end{description}

The most mathematical aspects of the pipeline are described in detail in the boxes 'Algorithm 1' and 'Algorithm 2', respectively. 

Specifically, step 1 in the  'Algorithm 1' box de-convolves all non-saturated images in order to reduce the effects of diffraction and diffusion. Then, steps from 2 through 6 construct a corresponding set of filtered Fourier transformed images (a Butterworth low-pass filter is also applied and, for each transformed image, just pixels where the filter values are greater than a threshold are retained). For each pixel, steps 7 through 11 construct two time sequences (for the real and imaginary parts of the Fourier transformed images) such that each time point contains the pixel intensity of the corresponding image in the sequence. Steps 12 and 14 use quadratic interpolation to infer the pixel values at those time points corresponding to the saturated images. Steps from 15 through 21 re-arrange these interpolation values to construct Fourier transform maps of the background at the saturation time points (in particular, steps 19 through 21 fill up the image regions cast away by the Butterworth filter by using the pixel contents of the Fourier transform of a reference unsaturated image). Steps 19 through 21 perform the inverse Fourier transform of the background images and finally step 24 projects such images onto the data space by means of the core diffusion PSF. We point out that working in the Fourier domain notably decreases the computational burden of the whole process, since most information in these data is concentrated in a limited region of small frequencies, where the interpolation is performed.

The 'Algorithm 2' box describes the pipeline routines that identify the primary saturation region with respect to the blooming one, apply EM to restore the flux in the primary saturation region and project this information into the data space in order to synthesize the de-saturated image. Specifically, steps from 2 through 9 realize  the segmentation of the saturated region by identifying the primary saturation region and determine the location of the diffraction fringes. This is obtained by thresholding the correlation between the diffraction PSF and the set of pixels contaminated by diffraction under the assumption that the whole saturation region were primary (the correlation map is computed by one iteration of EM). Step 10 through 12 restore the photon flux in the primary saturation region by numerically solving equation (\ref{model-equation}) via EM. Eventually, steps 13 through 25 provide the de-saturated image synthesis by glueing together the de-saturated information and the background.

\begin{algorithm}[b]
  \caption{background}
  \label{alg:Background}
  \begin{algorithmic}[1]
  \REQUIRE A dataset of AIA saturated and not saturated images corresponding to a fixed time interval. Let  $\{ s(p,j)$, $p=1,\cdots,N_{pixel}$, $j=0,\cdots,N_{sat}-1\}$ be the set of saturated images and $\{ ns(p,i)$, $p=1,\cdots,N_{pixel}$, $i=0,\cdots,N_{no\_sat}-1\}$ the set of  unsaturated images. Let $\mathcal{U}$ be the set of pixels in each image. Let $PSF$ and $PSF_C$ be the complete PSF and the core PSF of the AIA instrument respectively.  
  \ENSURE  A dataset of images $\{bg(\cdot,j)$, $j=0,\cdots,N_{sat}-1\}$ providing background estimation for saturated images based on information from not saturated ones.
  \STATE Deconvolve images $ ns(\cdot,\cdot)$ by using {\it PSF}: $ns^{\prime}(\cdot,\cdot) \leftarrow ns(\cdot,\cdot)$
    \FORALL{$i=0,\dots, N_{no\_sat}-1$}
    \STATE Compute the Fourier Transform of image $ns^{\prime}(\cdot,i)$: $\mathcal{F}ns^{\prime}(\cdot,i) \leftarrow ns^{\prime}(\cdot,i)$
    \STATE Apply a fourth-order Butterworth low-pass filter to image $\mathcal{F}ns^{\prime}(\cdot,i)$: $\mathcal{BF}ns^{\prime}(\cdot,i)\leftarrow \mathcal{F}ns^{\prime}(\cdot,i)$
    \ENDFOR
    \STATE Identify the set $\mathcal{S} \subset \mathcal{U}$ of pixels where the filter values are greater than a threshold value
     \FORALL{$p\in\mathcal{S}$}
      \FORALL{$i=0,\dots, N_{no\_sat}-1$}
   \STATE Fill entry $i$ of a vector {\it real\_orig}:  $real\_orig(i) \leftarrow\Re\left(\mathcal{BF}ns^{\prime}(p,i)\right)$     
    \STATE Fill entry $i$ of a vector {\it imag\_orig}:  $imag\_orig(i) \leftarrow \Im\left(\mathcal{BF}ns^{\prime}(p,i)\right)$ 
    \ENDFOR
   \STATE Apply interpolation to vector {\it real\_orig}:  $real\_interp \leftarrow real\_orig$
   \STATE Apply interpolation to vector {\it imag\_orig}:  $imag\_interp \leftarrow imag\_orig$
    \ENDFOR
     \FORALL{$j=0,\dots, N_{sat}-1$}
     \FORALL{$p\in\mathcal{S}$}
       \STATE  $\mathcal{F}bg^{\prime}(p,j) \leftarrow  {\rm complex}(real\_interp(j),imag\_interp(j))$  
     \ENDFOR
     \FORALL{$p\in\mathcal{U}\setminus S $}
     \STATE $\mathcal{F}bg^{\prime}(p,j) \leftarrow \mathcal{F}ns^{\prime}(p,0)$ 
     \ENDFOR
       \STATE Compute the Inverse Fourier Transform of image $\mathcal{F}bg^{\prime}(\cdot,j) $: $bg^{\prime}(\cdot,j) \leftarrow \mathcal{F}bg^{\prime}(\cdot,j) $ 
     \ENDFOR
     \STATE Convolve images $bg^{\prime}(\cdot,\cdot)$ by using $PSF_C$: $bg(\cdot,\cdot) \leftarrow bg^{\prime}(\cdot,\cdot)$
     \end{algorithmic}
\end{algorithm}

\begin{algorithm}[b]
  \caption{segmentation - EM reconstruction - image synthesis}
  \label{alg:Desaturation-step}
  \begin{algorithmic}[1]
  \REQUIRE A dataset of AIA saturated images $\{ s(p,j)$, $p=1,\cdots,N_{pixel}$, $j=0,\cdots,N_{sat}-1\}$ and  corresponding background estimations $\{ bg(p,j)$, $p=1,\cdots,N_{pixel}$, $j=0,\cdots,N_{sat}-1\}$. Let $\mathcal{U}$ be the set of pixels in each image. Let $PSF$, $PSF_C$ and $PSF_D$ be the complete PSF, the core PSF and the diffraction PSF of the AIA instrument respectively.  
  \ENSURE  A dataset of desaturated images $\{s_{desat}(\cdot,j)$, $j=0,\cdots,N_{sat}-1\}$.
   \FORALL{$j=0,\dots, N_{sat}-1$}
    \STATE Identify the set $\mathcal{S}^\prime \subset \mathcal{U}$ of saturated pixels
  \STATE Identify the set $\mathcal{NS}= \mathcal{U}\setminus \mathcal{S}^\prime$ of unsaturated pixels
  \STATE Convolve the indicator function $\mathbf{1}_{S^{\prime}}$  with $PSF_D$ to identify the set $\mathcal{F}^\prime \subset \mathcal{NS}$ of pixels where diffraction from pixels in $S^\prime$ virtually occurs: $\mathcal{F}^\prime \leftarrow \mathbf{1}_{S^{\prime}} * PSF_D$
  \STATE Compute the correlation $C^\prime$ between the image values in $\mathcal{F}^\prime$ and $PSF_D$ given the background estimation in $\mathcal{F}^\prime$
  \STATE Convolve $C^\prime$ with $PSF_C$ to obtain a correlation map $C$ in the image space: $C \leftarrow C^\prime*PSF_C$
  \STATE Threshold $C$ to identify the set $\mathcal{S} \subset \mathcal{S}^\prime$ of primary saturation pixels
  \STATE Identify the set $\mathcal{B} \subset \mathcal{S}^\prime$ of bloomed pixels: $\mathcal{B} \leftarrow \mathcal{S}^\prime \setminus \mathcal{S}$
    \STATE Convolve $\mathbf{1}_{S}$  with $PSF_D$ to identify the set $\mathcal{F} \subset \mathcal{F}^\prime$ of pixels where diffraction from pixels in $S$ actually occurs: $\mathcal{F} \leftarrow \mathbf{1}_{S} * PSF_D$
    \STATE Restrict $s(\cdot,j)$ image to $\mathcal{F}$:  $s_\mathcal{F}(j) \leftarrow s(p,j)$, $p\in\mathcal{F} $
       \STATE Restrict $bg(\cdot,j)$ image to $\mathcal{F}$:  $bg_\mathcal{F}(j) \leftarrow bg(p,j)$, $p\in\mathcal{F} $ 
    \STATE Apply $EM$ algorithm to solve the inverse problem given by Eq. (\ref{model-equation}) : $x_\mathcal{F}(j) \leftarrow EM(s_\mathcal{F}(j),bg_\mathcal{F}(j),PSF_D)$
         \STATE Restrict $bg(\cdot,j)$ image to $\mathcal{B}$:  $bg_\mathcal{B}(j) \leftarrow bg(p,j)$, $p\in\mathcal{B} $  
        \FORALL{$p\in\mathcal{U}$}
         \IF{$p\in\mathcal{F}$} 
         \STATE  $s_{desat}(p,j) \leftarrow [x_F(j)*PSF_C](p)$ 
         \ELSIF{$p\in\mathcal{B}$} 
         \STATE  $s_{desat}(p,j) \leftarrow bg_\mathcal{B}(p,j)$ 
         \ELSIF{$p\in\mathcal{F}$} 
         \STATE  $s_{desat}(p,j) \leftarrow s(p,j) -[x_F(j)*PSF_D](p)$
         \ELSE 
         \STATE$s_{desat}(p,j) \leftarrow s(p,j)$
          \ENDIF
        \ENDFOR
        \ENDFOR
      \end{algorithmic}
\end{algorithm}


\section{Numerical tests}
\begin{figure}
\begin{center}
\begin{tabular}{cc}
\includegraphics[width=8.cm]{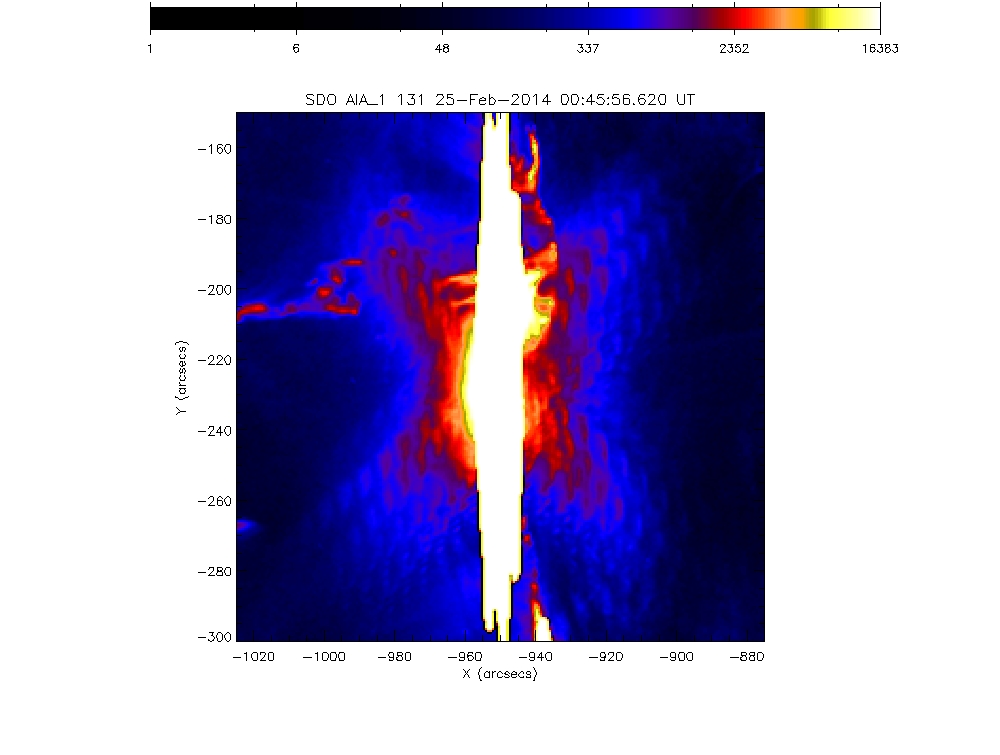} &
\includegraphics[width=8.cm]{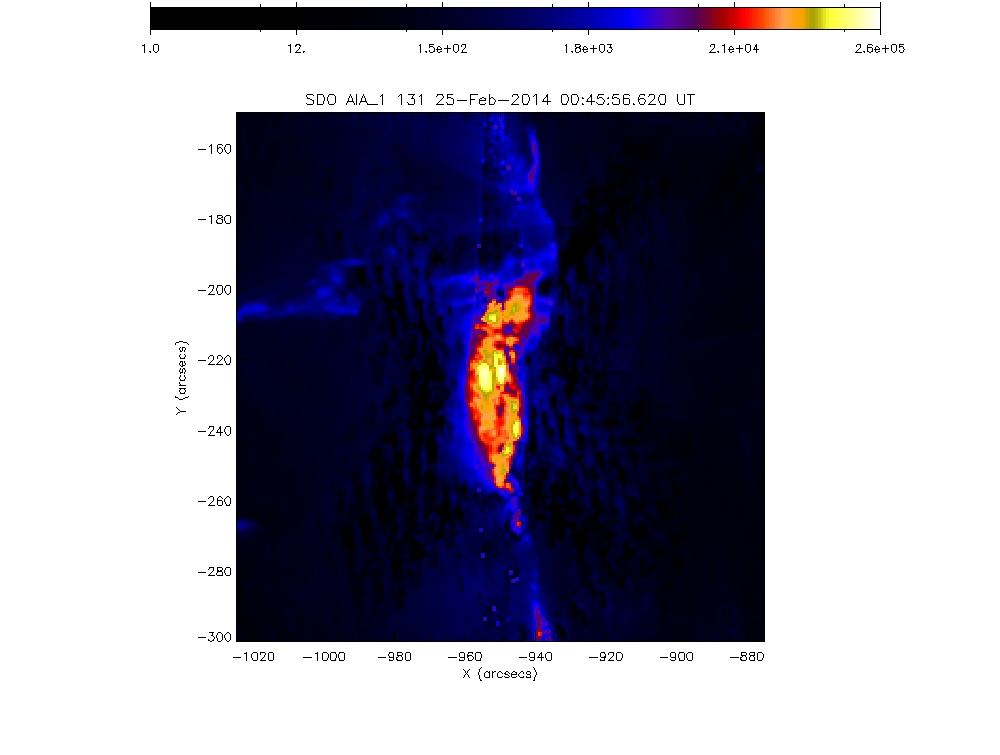} 
\end{tabular}
\caption{{\em{DESAT}} at work. Left panel: the same saturated image as in Figure \ref{fig:regions}. Right panel: the result of de-saturation.}
\label{fig:sat-de-sat}
\end{center}
\end{figure}

We tested {\em{DESAT}} in the case of the flaring event of February 25 2014, in the time interval between 00:45:00 UT and 00:47:00 UT, which presented nine saturated images with different saturation degree (i.e., different numbers of saturated pixels). An example of such images, corresponding to the time point 00:45:56, is represented in Figure \ref{fig:sat-de-sat}, which is the same one as in Figure \ref{fig:regions}; the right panel in the same figure shows the same image after the application of {\em{DESAT}}. In order to assess the relative and overall computational cost of the pipeline and its experimental reliability, we generated the saturated images according to three different sizes and tracked the computational cost of the 'segmentation', 'EM reconstruction' and 'image synthesis' elements of the pipeline.  The results of this test are in Table \ref{table:cinque} for the images with size $500 \times 500$, Table \ref{table:sette} for the images with size $750 \times 750$ and Table \ref{table:dieci} for the images with size $1000 \times 1000$, respectively. These results show that the biggest impact on the overall computational time of the pipeline comes from the EM step; this is expected, since this step requires several matrix times product computation. On the other hand the segmentation and image synthesis steps rather mildly depend on both the number of saturated/data pixel and the dimension of the image under de-saturation. 

As far as the reliability of the process is concerned, we computed the C-statistic values that measure the discrepancy between the observed data in the region of the diffraction fringes and the expectation values predicted by the reconstructed flux. The last column of the three Tables reports the averaged value of these C-statistic values and show that it significantly increases with the number of saturation points to cure. Further, Table \ref{table:c-stat} shows the behavior of the averaged C-statistic values when computed in correspondence of the fringe at positions with increasing distance from the image center (we reported here just the results concerning the $500 \times 500$ size, but the other cases behave coherently). Interestingly, the reliability of the de-saturation process seems not to be significantly influenced by the position of the reproduced diffraction fringes with respect to the de-saturated core.

\begin{table}
\begin{center}
\begin{tabular}{|c|c|c|c|c|c|c|c|}
\hline
time & $\#$ sat & $\#$ data & seg & EM (time) & EM (iter) & synthesis & C-stat \\
\hline
00:45:08 & 1304 & 112426 & 12.1 & 46.6 & 58 & 10.7& 34.8 \\
\hline
00:45:23 & 563 & 95712 & 12.7 & 14.6 & 53 & 10.8 & 2.3 \\
\hline
00:45:32 & 1459 & 110570 & 11.5 & 64.0 & 72 & 10.6 & 46.8 \\
\hline
00:45:48 & 23 & 16337 & 11.8 & 2.2& 37 & 10.4 & 3.9 \\
\hline
00:45:56 & 1592 & 121232 & 11.7 & 75.4 & 88 & 10.6 & 57.9 \\
\hline
00:46:12 & 26 & 20097 & 11.7 & 3.0 & 42 & 10.3 & 5.9 \\
\hline
00:46:20 & 1981 & 45691 & 11.8 & 64.4 & 71 & 10.1 & 26.4 \\
\hline
00:46: 36 & 147 & 45691 & 11.8 & 6.4 & 43 & 10.2 & 5.5 \\
\hline
00:46:44 & 2300 & 108488 & 11.4 & 71.4 & 67 & 10.2 & 21.6 \\
\hline
\end{tabular}
\caption{{\em{DESAT}} pipeline at work against images with $500 \times 500$ size. First column: recording time. Second column: number of saturated pixels. Third column: number of pixels in the diffraction fringe. Fourth column: computational time for the 'segmentation' step. Fifth column: computational time for the 'EM reconstruction' step. Sixth column: number of iteration for EM. Seventh column: computational time for the 'image synthesis' step. Eighth column: averaged C-statistic values computed in correspondence with the diffraction fringes.}
\label{table:cinque}
\end{center}
\end{table}

\begin{table}
\begin{center}
\begin{tabular}{|c|c|c|c|c|c|c|c|}
\hline
time & $\#$ sat & $\#$ data & seg & EM (time) & EM (iter) & synthesis & C-stat \\
\hline
00:45:08 & 1304 & 195097 & 15.5 & 280.1 & 55 & 10.7 & 39.0 \\
\hline
00:45:23 & 563 & 153760 & 15.8 & 25.8 & 50 & 108 & 2.0 \\
\hline
00:45:32 & 1459 & 193362 & 15.8 & 599.8 & 67 & 10.2 & 29.4 \\
\hline
00:45:48 & 23 & 20911 & 17.5 & 2.2 & 42 & 10.4 & 2.0 \\
\hline
00:45:56 & 1592 & 201723 & 18.7 & 502.0 & 92 & 10.3 & 36.1 \\
\hline
00:46:12 & 26 & 26327 & 15.9 & 20.0 & 37 & 10.4 & 20.0 \\
\hline
00:46:20 & 1981 & 205386 & 16.2 & 502.0 & 72 & 10.3 & 26.9 \\
\hline
00:46:36 & 147 & 62513 & 15.9 & 2.0 & 42 & 10.0 & 2.4 \\
\hline
00:46:44 & 2300 & 210492 & 16.3 & 368.1 & 73 & 10.4 & 13.4 \\
\hline
\end{tabular}
\caption{{\em{DESAT}} pipeline at work against images with $750 \times 750$ size. First column: recording time. Second column: number of saturated pixels. Third column: number of pixels in the diffraction fringe. Fourth column: computational time for the 'segmentation' step. Fifth column: computational time for the 'EM reconstruction' step. Sixth column: number of iteration for EM. Seventh column: computational time for the 'image synthesis' step. Eighth column: averaged C-statistic values computed in correspondence with the diffraction fringes.}
\label{table:sette}
\end{center}
\end{table}

\begin{table}
\begin{center}
\begin{tabular}{|c|c|c|c|c|c|c|c|}
\hline
time & $\#$ sat & $\#$ data & seg & EM (time) & EM (iter) & synthesis & C-stat \\
\hline
00:45:08 & 1304 & 237279 & 18.3 & 282.5 & 55 & 8.3 & 10.9 \\
\hline
00:45:23 & 563 & 178711 & 17.0 & 46.1 & 49 & 8.1 & 1.6 \\
\hline
00:45:32 & 1459 & 239203 & 16.3 & 356.0 & 69 & 8.4 & 36.1 \\
\hline
00:45:48 & 23 & 21312 & 16.3 & 2.2 & 26 & 8.2 & 1.6 \\
\hline
00:45:56 & 1592 & 246624 & 16.6 & 485.4 & 94 & 8.5 & 21.1 \\
\hline
00:46:12 & 26 & 28784 & 16.9 & 2.2 & 36 & 8.3 & 20.0 \\
\hline
00:46:20 & 1981 & 254602 & 16.6 & 401.8 & 75 & 8.5 & 21.8 \\
\hline
00:46:36 & 147 & 72714 & 18.3 & 5.5 & 38 & 8.8 & 1.7 \\
\hline
00:46:44 & 2300 & 263798 & 21.5 & 404.2 & 78 & 8.6 & 12.4 \\
\hline
\end{tabular}
\caption{{\em{DESAT}} pipeline at work against images with $1000 \times 1000$ size. First column: recording time. Second column: number of saturated pixels. Third column: number of pixels in the diffraction fringe. Fourth column: computational time for the 'segmentation' step. Fifth column: computational time for the 'EM reconstruction' step. Sixth column: number of iteration for EM. Seventh column: computational time for the 'image synthesis' step. Eighth column: averaged C-statistic values computed in correspondence with the diffraction fringes.}
\label{table:dieci}
\end{center}
\end{table}


\begin{table}
\begin{center}
\begin{tabular}{|c|c|c|c|}
\hline
25-Feb-2014& $R_1$ & $R_2$ &$R_3$\\
\hline
00:45:08 &2.21037 & 2.47712 & 2.37171 \\
\hline
00:45:23 & 0.991982 & 1.39406 & 1.16723 \\
\hline
00:45:32 & 2.36914 & 2.76948 & 2.54095 \\
\hline
00:45:48 &1.49663 & 2.11597 & 1.74499 \\
\hline
00:45:56 &7.87822 & 10.2115 & 9.05733 \\
\hline
00:46:12 &1.93773 & 1.68562 & 1.81033 \\
\hline
00:46:20 &6.91328 & 4.70813 & 5.73614 \\
\hline
00:46:36 &2.37601 & 1.69629 & 1.82247 \\
\hline
00:46:44 &4.32539 & 4.28034 & 4.01359 \\
\hline
\hline
\end{tabular}
\caption{Averaged C-statistic values computed in correspondence of the diffraction fringes in three image portions at increasing distance from the saturated core. $R_1$ corresponds to the region closest to the core (between $21$ arcsec and $60$ arcsec from the image center). $R_2$ corresponds to the intermediate region (between $60$ arcsec to $105$ arcsec). $R_3$ corresponds to the most peripheral region (between $105$ and $150$ arcsec).}
\label{table:c-stat}
\end{center}
\end{table}

\section{Conclusions}
{\em{DESAT}} is a first step to the full exploitation of {\em{SDO/AIA}} images: the availability of a fully automatic technological pipeline will make EUV data available also in the case of highly energetic flaring events (although saturation sometimes occurs even in the case of mildly energetic flares), and will realize this information improvement in the framework of a 'big data' approach, which is coherent with the {\em{SDO}} original concept. 

However, {\em{DESAT}} has still limitations and therefore it will probably pave the way to a series of actions that will ameliorate its performance at different levels. For example, there exists a reliability issue that should be addressed in {\em{DESAT}} and which is testified by still high C-statistic values, mainly in the case of strongly saturated data. To reduce such values, three aspects should be considered. First, all methods in {\em{DESAT}} strongly rely on the form of the PSF applied and therefore one should study the impact of the model of the PSF (particularly, its diffraction component) on the quality of de-saturation ({\em{DESAT}} is currently using the synthetic estimate of the diffraction PSF provided by SSW). Second, {\em{AIA}} filters provide data that are centered around a specific wavelength but that contain information coming from a range of wavelengths around the central one: accounting for this wavelength-dependent dispersion effect (which involves also the instrument PSF) should most likely help to reduce the C-statistic values. Third, the way {\em{DESAT}} is currently estimating the image background and therefore the way information is restored in the blooming region is most likely non-optimal and other strategies should be explored. 

Finally, at a more computational level, a significant improvement should come from the implementation of an automatic recipe to select the size of the image to de-saturate, in such a way that the corresponding input data will represent the optimal trade-off between the information content in the fringes and the computational burden.




\section*{References}

\bibliographystyle{model1a-num-names}
\bibliography{<your-bib-database>}

\begin{thebibliography}{999}
\bibitem{beba78} Bekefi, G., Barrett, A. H., 1978. Electromagnetic Vibrations, Waves, and Radiation, MIT Press.
\bibitem{bepi14} Benvenuto, F., Piana, M., 2014. Regularization of multiplicative iterative algorithms with non-negative constraint. Inverse Problems 30, 035012
\bibitem{beetal13} Benvenuto, F., Schwartz, R. A., Piana, M., Massone, A. M., 2013. Expectation Maximization for hard X-ray count modulation profiles. Astron. Astrophys. 555, A61.
\bibitem{gbetal06} Gnurek, S., Sylwester, J., Martens, P., 2006. The TRACE Telescope Point Spread Function for the 171  filter. Solar Phys. 239, 531-548.
\bibitem{leetal12} Lemen, J. R. et al, 2012. The Atmospheric Imaging Assembly (AIA) on the Solar Dynamics Observatory (SDO). Solar Phys. 275, 207.
\bibitem{makl97} Martinez, P., Klotz, A., 1997. A practical Guide to CCD Astronomy, Cambridge University Press.
\bibitem{sctopi14} Schwartz, R. A., Torre, G., Piana, M., 2014. Systematic de-saturation of images form the Atmospheric Imaging Assembly in the Solar Dynamics Observatory. Astrophys. J. Lett. 793, L23.
\bibitem{shva82} Shepp, L. A., Vari, Y., 1982. Maximum likelihood reconstruction for emission tomography. IEEE Trans. Med. Imaging 1, 113.
\bibitem{soetal00} Socker, D. G., Howard, R.A., Korendyke, C.M., Simnett, G.M., Webb D. F., 2000. NASA Solar Terrestrial Relations Observatory (STEREO) mission heliospheric imager. Proc. SPIE 4139, 284.
\bibitem{toetal15} Torre, G., Schwartz, R.A., Benvenuto, F., Massone, A. M., Piana, M., 2015. Inverse diffraction for the Atmospheric Imaging Assembly in the Solar Dynamics Observatory. arXiv: 1501.07805.

\end{thebibliography}







\end{document}